\documentclass[%
 reprint,
 superscriptaddress,
 amsmath,
 amssymb,
 aps,
prl,
]{revtex4-2}

\usepackage{graphicx}
\usepackage{dcolumn}
\usepackage{bm}
\usepackage{hyperref}
\usepackage[mathlines]{lineno}
\usepackage{siunitx}
\hypersetup{colorlinks=true, urlcolor=blue, citecolor=blue, linkcolor=blue,}
\usepackage{capt-of}

\usepackage{nasa}
\usepackage{booktabs}
\usepackage{multirow}
\usepackage{enumitem}
\usepackage{xcolor}

\usepackage{comment}

\begin{document}

\preprint{APS/123-QED}

\title{Indication of Electron Antineutrinos from the Diffuse Supernova Neutrino Background}



\newcommand{\AFFicrr}{\affiliation{Kamioka Observatory, Institute for Cosmic Ray Research, University of Tokyo, Kamioka, Gifu 506-1205, Japan}}
\newcommand{\AFFkashiwa}{\affiliation{Research Center for Cosmic Neutrinos, Institute for Cosmic Ray Research, University of Tokyo, Kashiwa, Chiba 277-8582, Japan}}
\newcommand{\AFFipmu}{\affiliation{Kavli Institute for the Physics and
Mathematics of the Universe (WPI), The University of Tokyo Institutes for Advanced Study,
University of Tokyo, Kashiwa, Chiba 277-8583, Japan }}
\newcommand{\AFFmad}{\affiliation{Department of Theoretical Physics, University Autonoma Madrid, 28049 Madrid, Spain}}
\newcommand{\AFFubc}{\affiliation{Department of Physics and Astronomy, University of British Columbia, Vancouver, BC, V6T1Z4, Canada}}
\newcommand{\AFFbu}{\affiliation{Department of Physics, Boston University, Boston, MA 02215, USA}}
\newcommand{\AFFuci}{\affiliation{Department of Physics and Astronomy, University of California, Irvine, Irvine, CA 92697-4575, USA }}
\newcommand{\AFFcsu}{\affiliation{Department of Physics, California State University, Dominguez Hills, Carson, CA 90747, USA}}
\newcommand{\AFFcnm}{\affiliation{Institute for Universe and Elementary Particles, Chonnam National University, Gwangju 61186, Korea}}
\newcommand{\AFFduke}{\affiliation{Department of Physics, Duke University, Durham NC 27708, USA}}
\newcommand{\AFFgifu}{\affiliation{Department of Physics, Gifu University, Gifu, Gifu 501-1193, Japan}}
\newcommand{\AFFgist}{\affiliation{GIST College, Gwangju Institute of Science and Technology, Gwangju 500-712, Korea}}
\newcommand{\AFFuh}{\affiliation{Department of Physics and Astronomy, University of Hawaii, Honolulu, HI 96822, USA}}
\newcommand{\AFFicl}{\affiliation{Department of Physics, Imperial College London , London, SW7 2AZ, United Kingdom }}
\newcommand{\AFFkek}{\affiliation{High Energy Accelerator Research Organization (KEK), Tsukuba, Ibaraki 305-0801, Japan }}
\newcommand{\AFFkobe}{\affiliation{Department of Physics, Kobe University, Kobe, Hyogo 657-8501, Japan}}
\newcommand{\AFFkyoto}{\affiliation{Department of Physics, Kyoto University, Kyoto, Kyoto 606-8502, Japan}}
\newcommand{\AFFliv}{\affiliation{Department of Physics, University of Liverpool, Liverpool, L69 7ZE, United Kingdom}}
\newcommand{\AFFmiyagi}{\affiliation{Department of Physics, Miyagi University of Education, Sendai, Miyagi 980-0845, Japan}}
\newcommand{\AFFnagoya}{\affiliation{Institute for Space-Earth Environmental Research, Nagoya University, Nagoya, Aichi 464-8602, Japan}}
\newcommand{\AFFkmi}{\affiliation{Kobayashi-Maskawa Institute for the Origin of Particles and the Universe, Nagoya University, Nagoya, Aichi 464-8602, Japan}}
\newcommand{\AFFpol}{\affiliation{National Centre For Nuclear Research, 02-093 Warsaw, Poland}}
\newcommand{\AFFsuny}{\affiliation{Department of Physics and Astronomy, State University of New York at Stony Brook, NY 11794-3800, USA}}
\newcommand{\AFFokayama}{\affiliation{Department of Physics, Okayama University, Okayama, Okayama 700-8530, Japan }}
\newcommand{\AFFosaka}{\affiliation{Department of Physics, Osaka University, Toyonaka, Osaka 560-0043, Japan}}
\newcommand{\AFFox}{\affiliation{Department of Physics, Oxford University, Oxford, OX1 3PU, United Kingdom}}
\newcommand{\AFFqmul}{\affiliation{School of Physics and Astronomy, Queen Mary University of London, London, E1 4NS, United Kingdom}}
\newcommand{\AFFregina}{\affiliation{Department of Physics, University of Regina, 3737 Wascana Parkway, Regina, SK, S4SOA2, Canada}}
\newcommand{\AFFseoul}{\affiliation{Department of Physics and Astronomy, Seoul National University, Seoul 151-742, Korea}}
\newcommand{\AFFsheff}{\affiliation{School of Mathematical and Physical Sciences, University of Sheffield, S3 7RH, Sheffield, United Kingdom}}
\newcommand{\AFFshizuokasc}{\affiliation{Department of Informatics in
Social Welfare, Shizuoka University of Welfare, Yaizu, Shizuoka, 425-8611, Japan}}
\newcommand{\AFFstfc}{\affiliation{STFC, Rutherford Appleton Laboratory, Harwell Oxford, and Daresbury Laboratory, Warrington, OX11 0QX, United Kingdom}}
\newcommand{\AFFskk}{\affiliation{Department of Physics, Sungkyunkwan University, Suwon 440-746, Korea}}
\newcommand{\AFFtodai}{\affiliation{Department of Physics, University of Tokyo, Bunkyo, Tokyo 113-0033, Japan }}
\newcommand{\AFFtit}{\affiliation{Department of Physics, Institute of Science Tokyo, Meguro, Tokyo 152-8551, Japan }}
\newcommand{\AFFtus}{\affiliation{Department of Physics and Astronomy, Faculty of Science and Technology, Tokyo University of Science, Noda, Chiba 278-8510, Japan }}
\newcommand{\AFFtriumf}{\affiliation{TRIUMF, 4004 Wesbrook Mall, Vancouver, BC, V6T2A3, Canada }}
\newcommand{\AFFtokai}{\affiliation{Department of Physics, Tokai University, Hiratsuka, Kanagawa 259-1292, Japan}}
\newcommand{\AFFtsinghua}{\affiliation{Department of Engineering Physics, Tsinghua University, Beijing, 100084, China}}
\newcommand{\AFFynu}{\affiliation{Department of Physics, Yokohama National University, Yokohama, Kanagawa, 240-8501, Japan}}
\newcommand{\AFFllr}{\affiliation{Ecole Polytechnique, IN2P3-CNRS, Laboratoire Leprince-Ringuet, F-91120 Palaiseau, France }}
\newcommand{\AFFbari}{\affiliation{ Dipartimento Interuniversitario di Fisica, INFN Sezione di Bari and Universit\`a e Politecnico di Bari, I-70125, Bari, Italy}}
\newcommand{\AFFnapoli}{\affiliation{Dipartimento di Fisica, INFN Sezione di Napoli and Universit\`a di Napoli, I-80126, Napoli, Italy}}
\newcommand{\AFFroma}{\affiliation{INFN Sezione di Roma and Universit\`a di Roma ``La Sapienza'', I-00185, Roma, Italy}}
\newcommand{\AFFpadova}{\affiliation{Dipartimento di Fisica, INFN Sezione di Padova and Universit\`a di Padova, I-35131, Padova, Italy}}
\newcommand{\AFFkeio}{\affiliation{Department of Physics, Keio University, Yokohama, Kanagawa, 223-8522, Japan}}
\newcommand{\AFFwinnipeg}{\affiliation{Department of Physics, University of Winnipeg, MB R3J 3L8, Canada }}
\newcommand{\AFFkcl}{\affiliation{Department of Physics, King's College London, London, WC2R 2LS, UK }}
\newcommand{\AFFwarwick}{\affiliation{Department of Physics, University of Warwick, Coventry, CV4 7AL, UK }}
\newcommand{\AFFral}{\affiliation{Rutherford Appleton Laboratory, Harwell, Oxford, OX11 0QX, UK }}
\newcommand{\AFFwu}{\affiliation{Faculty of Physics, University of Warsaw, Warsaw, 02-093, Poland }}
\newcommand{\AFFbcit}{\affiliation{Department of Physics, British Columbia Institute of Technology, Burnaby, BC, V5G 3H2, Canada }}
\newcommand{\AFFtohoku}{\affiliation{Department of Physics, Faculty of Science, Tohoku University, Sendai, Miyagi, 980-8578, Japan }}
\newcommand{\AFFicise}{\affiliation{Institute For Interdisciplinary Research in Science and Education, ICISE, Quy Nhon, 55121, Vietnam }}
\newcommand{\AFFilance}{\affiliation{ILANCE, CNRS - University of Tokyo International Research Laboratory, Kashiwa, Chiba 277-8582, Japan}}
\newcommand{\AFFibs}{\affiliation{Center for Underground Physics, Institute for Basic Science (IBS), Daejeon, 34126, Korea}}
\newcommand{\AFFglasgow}{\affiliation{School of Physics and Astronomy, University of Glasgow, Glasgow, Scotland, G12 8QQ, United Kingdom}}
\newcommand{\AFFoecu}{\affiliation{Media Communication Center, Osaka Electro-Communication University, Neyagawa, Osaka, 572-8530, Japan}}
\newcommand{\AFFminn}{\affiliation{School of Physics and Astronomy, University of Minnesota, Minneapolis, MN  55455, USA}}
\newcommand{\AFFsilesia}{\affiliation{August Che\l{}kowski Institute of Physics, University of Silesia in Katowice, 75 Pu\l{}ku Piechoty 1, 41-500 Chorz\'{o}w, Poland}}
\newcommand{\AFFtoyama}{\affiliation{Faculty of Science, University of Toyama, Toyama City, Toyama 930-8555, Japan}}
\newcommand{\AFFbmcc}{\affiliation{Science Department, Borough of Manhattan Community College / City University of New York, New York, New York, 1007, USA.}}
\newcommand{\AFFnumazu}{\affiliation{National Institute of Technology, Numazu College, Numazu, Shizuoka  410-8501, Japan}}
\newcommand{\AFFniihama}{\affiliation{National Institute of Technology, Niihama College, Niihama, Ehime  792-8580, Japan}}
\newcommand{\AFFucas}{\affiliation{School of Physical Sciences, University of Chinese Academy of Sciences, Beijing 101408, China}}

\AFFicrr
\AFFkashiwa
\AFFmad
\AFFbmcc
\AFFbu
\AFFbcit
\AFFuci
\AFFcsu
\AFFucas
\AFFcnm
\AFFduke
\AFFllr
\AFFgifu
\AFFgist
\AFFglasgow
\AFFuh
\AFFibs
\AFFicise
\AFFicl
\AFFbari
\AFFnapoli
\AFFpadova
\AFFroma
\AFFilance
\AFFkeio
\AFFkek
\AFFkcl
\AFFkobe
\AFFkyoto
\AFFliv
\AFFminn
\AFFmiyagi
\AFFnagoya
\AFFkmi
\AFFpol
\AFFniihama
\AFFnumazu
\AFFsuny
\AFFokayama
\AFFoecu
\AFFox
\AFFral
\AFFseoul
\AFFsheff
\AFFshizuokasc
\AFFsilesia
\AFFstfc
\AFFskk
\AFFtohoku
\AFFtodai
\AFFipmu
\AFFtit
\AFFtus
\AFFtoyama
\AFFtriumf
\AFFtsinghua
\AFFwu
\AFFwarwick
\AFFwinnipeg
\AFFynu

\author{K.~Abe}
\AFFicrr
\AFFipmu
\author{Y.~Asaoka}
\AFFicrr
\AFFipmu
\author{M.~Harada}
\email{Corresponding author: mharada@km.icrr.u-tokyo.ac.jp}
\AFFicrr
\author{Y.~Hayato}
\AFFicrr
\AFFipmu
\author{K.~Hiraide}
\AFFicrr
\AFFipmu
\author{T.~H.~Hung}
\AFFicrr
\author{K.~Ieki}
\author{M.~Ikeda}
\AFFicrr
\AFFipmu
\author{J.~Kameda}
\AFFicrr
\AFFipmu
\author{Y.~Kataoka}
\AFFicrr
\AFFipmu
% ...


\collaboration{Super-Kamiokande Collaboration}
\noaffiliation

\date{\today}

\begin{abstract}
We report an indication of $\bar{\nu}_e$ from the diffuse supernova neutrino background (DSNB) based on a spectral fit analysis of a 5003.0-day dataset in Super-Kamiokande. Key improvements over previous searches include enhanced rejection of atmospheric $\nu$ backgrounds, updated evaluations of systematic uncertainties, and a refined spectral fitting method. Using a representative DSNB model, the best-fit flux was determined to be $3.6^{+1.6}_{-1.5}~\mathrm{cm^{-2}\,s^{-1}}$ for $E_\nu>13.3$~MeV, corresponding to a $2.6\sigma$ tension with a null hypothesis ($99.5\%$ C.L., one-sided). Similar significances were obtained with signal models of varying spectral shapes. 

\end{abstract}

\maketitle

 
\textit{Introduction---}Since the historic detection of 24 neutrinos from SN~1987A~\cite{1987JETPL..45..589A,1987PhRvL..58.1494B,1987PhRvL..58.1490H}, a further detection of supernova neutrinos has been a central goal of neutrino astronomy. Core-collapse supernovae are expected to release on the order of $10^{53}$ erg of energy, approximately 99\% of which is in the form of neutrinos~(e.g., see~\cite{2006RPPh...69..971K,2012ARNPS..62..407J,2013RvMP...85..245B,2018SSRv..214...31T,2020LRCA....6....3M,2021Natur.589...29B,2024RvMP...96b5004V}). Detecting bursts of neutrinos from nearby supernovae is important, but the burst rate is only a few per century~\cite{1994ApJS...92..487T}. Integrating over the observable universe, however, core collapses occur a few times per second~\cite{1994ApJS...92..487T,2006Natur.439...45D, 2025A&A...693A.306F}. The diffuse supernova neutrino background (DSNB) is the steady, isotropic flux of neutrinos and antineutrinos accumulated from all such collapses since the beginning of star formation~\cite{2010ARNPS..60..439B,2023PJAB...99..460A}. A measurement of its spectrum would offer unique insight into not only the core collapse mechanism, but also the fraction of collapses that form black holes and galactic chemical evolution. Many theoretical predictions have been made~\cite{1995APh.....3..367T,1997APh.....7..137H,2003APh....18..307A,2003PhLB..559..113A,2004NJPh....6..170A,2009PhRvD..79h3013H,2009PhRvL.102w1101L,2013PhRvD..88h3012N,2015ApJ...804...75N,2018MNRAS.475.1363H,2021PhRvD.103d3003H,2021ApJ...909..169K,2021JCAP...05..011T,2022PhRvD.106d3026E,2022ApJ...937...30A,2022MNRAS.517.2471Z,2023ApJ...953..151A,2023PhRvD.107b3017I,2024PhRvD.109b3024E,2024ApJ...975...71N,2024PhRvD.110j3029M,2026PhRvD.113f3044L} and experimental searches have been conducted in water Cherenkov and liquid scintillator detectors~\cite{2003PhRvL..90f1101M,2012PhRvD..85e2007B,2012ApJ...745..193G,2015APh....60...41Z,2020PhRvD.102f2006A,2021PhRvD.104l2002A,2021APh...12502509A,2022ApJ...925...14A,2023ApJ...951L..27H,2026ApJ..1005..101A,2026arXiv260629381C}, but no significant excess has been observed so far.

All neutrino flavors are emitted in a core collapse, but most DSNB searches target $\bar{\nu}_e$ via inverse $\beta$ decay (IBD) on free protons ($\bar{\nu}_e + p \rightarrow e^+ + n$, $E_e = E_\nu - 1.3\,\mathrm{MeV}$)~\cite{2003PhLB..564...42S}, since in the relevant energy range for detection, roughly $E_e = 10\text{--}30\,\mathrm{MeV}$, the IBD cross section is 2 orders of magnitude larger than that of the subdominant $\nu$-electron elastic scattering~\cite{2012RvMP...84.1307F}. Within this energy range, $\beta$ decays of radioisotopes produced by cosmic-ray muon spallation and atmospheric $\nu$ interactions are dominant backgrounds. The expected DSNB signal is only a few events per year~\cite{2010ARNPS..60..439B}, making the efficient and reliable tagging of the delayed neutron in coincidence with the prompt positron an important tool to suppress some backgrounds and to enhance the sensitivity of the search~\cite{2004PhRvL..93q1101B}.

In this Letter, we present a search for DSNB $\bar{\nu}_e$ using neutron tagging in Super-Kamiokande (SK) with data collected between October 2008 and March 2026, yielding a total livetime of 5003.0 days. We first perform a DSNB model-independent analysis, comparing the number of observed events with the background expectation in the energy range $13.3 < E_{\nu} < 31.3$~MeV. We then perform a spectral fit analysis, in which the observed spectrum is fitted with spectra of the signal and six background components over the wider range $13.3 < E_{\nu} < 81.3$~MeV. From the latter, our primary analysis, we report the first indication ($2.6\sigma$) of the DSNB $\bar{\nu}_e$ signal.

SK is a water Cherenkov detector, with a 22.5-kton analysis fiducial volume, located in the Kamioka mine in Gifu, Japan~\cite{2003NIMPA.501..418F}. The data reported here span six operational phases: SK-IV, V, VI, VII, VII.5, and VIII. Neutron tagging became possible starting in SK-IV (2970.1 days) following an electronics upgrade~\cite{2010ITNS...57..428Y}. Between SK-IV and SK-V, detector refurbishment work was carried out in preparation for dissolving gadolinium sulfate into SK's pure water. Then SK-V (379.2 days) operated as the last pure-water phase. Starting in SK-VI (552.2 days), 13 tons of $\mathrm{Gd_2(SO_4)_3 \cdot 8H_2O}$ was dissolved~\cite{2022NIMPA102766248A} to enhance the neutron tagging efficiency~\cite{2004PhRvL..93q1101B}. In SK-VII (404.0 days), an additional 26.1 tons of the Gd compound was dissolved, raising the Gd mass concentration from $0.011\%$ to $0.033\%$~\cite{2024NIMPA106569480A}, and the neutron capture fraction on Gd from $50$\% to $75$\%. In October 2023, a failure occurred in the geomagnetic compensation coils arranged around the inner detector. Accordingly, in SK-VII.5 (191.8 days), photomultiplier tubes (PMTs) affected by the failure operated with reduced light collection efficiency. Following repairs to the compensation coil system, SK-VIII (505.7 days) was started and continues today. 

\textit{Methodology---}The IBD interactions of the DSNB signal are simulated using the supernova event generator SKSNSim~\cite{2024ApJ...965...91N}. Atmospheric $\nu$ backgrounds above $100$~MeV are simulated using the HKKM 2011 flux~\cite{2007PhRvD..75d3006H,2011PhRvD..83l3001H} as input to the neutrino event generator NEUT 5.9.2~\cite{2009AcPPB..40.2477H,2021EPJST.230.4469H}. Quasielastic (QE) interactions, dominant in the relevant search energy range, are described by the spectral function model~\cite{1994NuPhA.579..493B, 2005PhRvD..72e3005B, 2025PhRvD.111c3006A}, and nuclear deexcitation by NucDeEx~\cite{2023EPJA...59..131K, 2024PhRvD.109c6009A,2026EPJC...86..148H}. The detector response is modeled by SKG4, a Geant4-based SK detector simulation~\cite{2003NIMPA.506..250A, 2006ITNS...53..270A, 2016NIMPA.835..186A, 2020JPhCS1468a2255H}, using the INCL cascade model coupled with NucDeEx~\cite{2025PhRvD.112k2013A}. All Monte Carlo (MC) samples and data are reconstructed using a dedicated low-energy event fitter~\cite{2024PhRvD.109i2001A} and run through the event selection steps described below. The total energy, $E_e$, of the prompt DSNB interaction candidates is reconstructed assuming all light is emitted from a single electron or positron. The reconstructed energy, $E_{\rm rec}$, is defined as the corresponding kinetic energy of that electron or positron. For IBD candidates the relation with neutrino energy is $E_\nu \simeq E_{\rm rec} + 1.8$~MeV.

Positron candidate events above $\sim$6 MeV are collected by a dedicated trigger, whose threshold (hit PMTs in 200~ns) was 70 until September 2011 and 58 thereafter in SK-IV, 60--70 in SK-V, and 60 from SK-VI onward. The prompt event window of $35$~$\mu$s after the trigger time is followed by another $500$~$\mu$s where all PMT hits are recorded. The full $535$~$\mu$s is sufficient for the coincidence detection of the prompt positron and delayed neutron, since the expected neutron capture times range from ${\sim}200$~$\mu$s in pure water to ${\sim}60$~$\mu$s in 0.033\% Gd water. The event selection then proceeds by application of noise cuts~\cite{2021PhRvD.104l2002A,2026ApJ..1005..101A}, atmospheric $\nu$ and spallation background cuts, and neutron tagging. 

Atmospheric $\nu$ interactions form major backgrounds to the DSNB signal. Neutral-current (NC) interactions resulting in nuclear deexcitation $\gamma$ rays, accompanied by a knocked-out neutron, mimic IBD~\cite{2019PhRvD..99c2005W,2019PhRvD.100k2009A,2020.PhDThesis.YAshida.KyotoU,2024PhRvD.109a1101S}. At higher energies, charged-current (CC) interactions of $\nu_\mu$ and $\bar{\nu}_\mu$ resulting in a decay electron together with a neutron can also mimic IBD~\cite{2021PhRvD.104l2002A}. These backgrounds are separated from signal by the pattern and timing of the prompt event's PMT hits: deexcitation $\gamma$ rays produce hit patterns more isotropic than that of a single positron, while muons decay with a lifetime of 2.2~$\mu$s, producing an additional cluster of hits within the prompt event window. Two improvements to these rejection methods were made in the present analysis. 

First, a machine learning algorithm was developed to target the NC interactions~\cite{2026.PhDThesis.LFeng.KyotoU}. Its inputs are the reconstructed Cherenkov angle ($\theta_{\mathrm{C}}$)~\cite{2003PhRvL..90f1101M}, which quantifies the opening angle of the Cherenkov cone; multiple scattering goodness ($\mathrm{MSG}$)~\cite{2012PhRvD..85e2007B, 2016PhRvD..94e2010A, 2024PhRvD.109i2001A}, which quantifies the fuzziness of the Cherenkov ring from a low-energy electron; and the maximum number of hits in a sliding 50 ns window ($\mathrm{n50}$). The first two assume a single Cherenkov cone, which does not hold for NC interactions with multiple $\gamma$ rays, and hence multiple Compton electrons. If the $\gamma$ rays go in different directions, they reconstruct with unphysical opening angles ($\gg 41^\circ$) and artificially small MSG values. Even if the $\gamma$ rays all align, MSG will be lower, since each Compton electron has lower energy and therefore increased multiple Coulomb scattering. Compared with previous cuts~\cite{2026ApJ..1005..101A, 2026.PhDThesis.LFeng.KyotoU}, this technique increases the IBD selection efficiency from 79\% to 89\% and halves the remaining NCQE fraction, from 10\% to 5\%.

To reduce CC interactions, previous analyses required no decay electron candidates within the prompt event window~\cite{2021PhRvD.104l2002A,2023ApJ...951L..27H,2026ApJ..1005..101A}. In the Gd-water phases, this rejected IBD events whose neutron captures were misidentified as decay electrons, decreasing the signal efficiency by $\sim$10\%. We instead separate decay electrons from Gd neutron captures by finding the largest number of time-of-flight corrected hits in a sliding 15~ns window after the prompt peak, and using the ratio of this hit cluster's time to its number of PMT hits. Decay electrons appear within a few microseconds of the prompt peak and produce more hits, while Gd neutron captures occur tens of microseconds later with fewer hits.

Cosmic-ray muon spallation with oxygen nuclei produces $\beta$-unstable isotopes at a rate up to $10^6$ times the predicted DSNB signal below about 20 MeV~\cite{2021PhRvD.104l2002A}. Most decay by $\beta$ emission alone and are removed by the requirement of a delayed neutron, entering the sample only in conjunction with a falsely tagged neutron. What remains is further reduced by correlating candidate events with the timing, reconstructed track, and energy deposition profile along the track of preceding muons~\cite{2012PhRvD..85e2007B, 2024PhRvD.110c2003L}. These spallation cuts and neutron tagging were optimized simultaneously to suppress the resulting accidental coincidence background. In contrast, $^9\mathrm{Li}$ undergoes $\beta+n$ decay~\cite{2014PhRvC..89d5801L,2015PhRvD..91j5005L,2015PhRvD..92j5033L,2025PhRvD.111b3014N}, exactly reproducing the IBD signature. Given its $0.178$~s half-life, $^9\mathrm{Li}$ is difficult to efficiently remove with current spallation cuts. We therefore estimate its surviving contribution by normalizing the $\beta$ decay spectrum given in~\cite{2021PhRvD.104l2002A} to the $^9\mathrm{Li}$ production yield of $0.86 \text{ kton}^{-1} \cdot \text{day}^{-1}$ from SK's spallation measurement~\cite{2016PhRvD..93a2004Z}. Other $\beta+n$ emitters in the relevant energy range have small yields (e.g., $^{11}$Li) and are negligible. 

Delayed neutrons are identified using a neural network~\cite{2026ApJ..1005..101A, 2023.PhDThesis.SHan.TokyoU, 2026.PhDThesis.BZhang.TsinghuaU} trained separately for each phase using common input variables. The selection threshold is set per energy range in each phase to suppress the accidental coincidences. It is tightest at the lowest energies, where the number of events before neutron tagging is largest, and is gradually loosened above $E_{\rm rec} = 15.5$~MeV, where accidental coincidences are subdominant. The resulting probability of falsely tagging a delayed neutron is $10^{-4}$ to $10^{-3}$ in all phases at the lowest energies, rising to $10^{-1}$ in the pure-water phases at the highest. The corresponding neutron tagging efficiencies are 10--30\% in the pure-water phases, 40--50\% in the 0.011\% Gd-water phase, and 60--70\% in the 0.033\% Gd-water phases.

Systematic uncertainties for the atmospheric $\nu$ backgrounds arise from the interaction rate (flux times cross section) and the final-state kinematics ($\gamma$ ray and neutron multiplicities). For CC interactions, 20\% is assigned to the flux \cite{2011PhRvD..83l3001H} and 30\% to the cross section~\cite{2012RvMP...84.1307F,2014NJPh...16g5015A,2018PhRvD..97a2001A,2025PhRvD.112k2020A}. For NC interactions, the selection efficiency depends strongly on the $\gamma$ ray multiplicity, which is uncertain owing to final-state interactions inside the nucleus and secondary interactions in the detector~\cite{2024PhRvC.109a4620A,2024PTEP.2024k3D01T}. We therefore compared data and MC event rates after all cuts except neutron tagging using an independent T2K beam sample with the same final state topology~\cite{2019PhRvD.100k2009A,2025PhRvD.112c2003A,2026.PhDThesis.LFeng.KyotoU}, giving a 45\% NC rate uncertainty, improved from the previous 70\%~\cite{2021PhRvD.104l2002A,2026ApJ..1005..101A}. The neutron multiplicity uncertainty was estimated for CC and NC events by shaping the MC multiplicity distribution to match the mean value~\cite{2020.PhDThesis.YAshida.KyotoU,2021PhRvD.104l2002A} of SK's measurement~\cite{2025PhRvD.112a2004H}, reducing the uncertainty from the previous 30--40\%~\cite{2019.PhDThesis.RAkutsu.UTokyo,2021PhRvD.104l2002A,2026ApJ..1005..101A} to 10--30\%. 

Systematic uncertainties for the spallation $^9\mathrm{Li}$ background are taken on the isotope yield and the remaining rate after cuts. For the yield, $22\%$ is assigned based on~\cite{2016PhRvD..93a2004Z}. Previous analyses assigned a conservative $50\%$ error to the remaining rate, estimated from the $^9\mathrm{Li}$ half-life alone and assuming the same spatial correlation with the parent muon as other isotopes~\cite{2021PhRvD.104l2002A}. We instead evaluate this correlation directly by simulating $^9\mathrm{Li}$ production with SKG4 using the Bertini cascade model~\cite{2016NIMPA.835..186A}. The simulated distance distribution to the parent muon is consistent with a $^9\mathrm{Li}$-dominant data sample selected with loose neutron tagging and spallation cuts applied to events below $E_{\rm rec} = 15.5$~MeV. Comparing estimated and true $^9\mathrm{Li}$ efficiencies in simulation, the difference in spatial correlation was found to be subdominant to that in temporal correlation, reducing the remaining rate error to $30\%$.

The neutron tagging systematic uncertainty was evaluated with an americium-beryllium calibration source~\cite{2009APh....31..320S}. The data--MC difference in tagging efficiency was found to be $10\%$ for the pure-water phases, $3\%$ for the 0.011\% Gd-water phase, and $5\%$ for the 0.033\% Gd-water phases. Table~\ref{tab:tab_1} summarizes all systematics.

\textit{Statistical Analysis---}The data were divided into six analysis regions based on the reconstructed Cherenkov angle of the positron candidate ($\theta_{\mathrm{C}} \in [20^\circ, 38^\circ]$, $\theta_{\mathrm{C}} \in [38^\circ, 53^\circ]$, or $\theta_{\mathrm{C}} \in [78^\circ, 90^\circ]$) and the number of tagged neutrons ($N_{\text{n}} \neq 1$ or $N_{\text{n}} = 1$). IBD positrons mostly populate the mid-$\theta_{\mathrm{C}}$ band ($\theta_{\mathrm{C}} \in [38^\circ, 53^\circ]$, the signal region) in either $N_{\text{n}} = 1$ or $N_{\text{n}} \neq 1$ subregions, depending on the neutron tagging efficiencies of each phase. The remaining four regions are background-dominated sidebands.

In the DSNB model-independent analysis, events in the signal region with $N_{\text{n}} = 1$ are compared with the expected background. Data at $E_{\rm rec} \in [29.5, 79.5]$~MeV were used to scale the atmospheric $\nu$ non-NCQE event component at lower energies. The resulting energy spectra are shown in Figure~\ref{fig:binned_final_spectra}. The background-only $p$-value is obtained by comparing the observed count against a distribution built from pseudoexperiments, in which each background category is varied within its systematic uncertainty assuming Gaussian distributions~\cite{2021PhRvD.104l2002A,2026ApJ..1005..101A}. Applying this to the summed counts over $E_{\rm rec} \in [11.5, 29.5]$~MeV, we observe a $1.8\sigma$ tension with the background-only hypothesis ($p$-value 0.036). 

The spectral fit analysis uses all six analysis regions. To observe the DSNB over as wide an energy range as possible, the $N_{\rm n} = 1$ threshold was lowered from the previous analysis's $E_{\rm rec} = 15.5$~MeV to 11.5~MeV. This adds spallation $^9$Li (whose remaining rate is now constrained by simulation validated with data) and accidental coincidences as fit components. In the $N_{\rm n} \neq 1$ subregion, spallation $\beta$ decays without an accompanying neutron are not suppressed by neutron tagging. Previous searches~\cite{2021PhRvD.104l2002A,2026ApJ..1005..101A} modeled this contribution in $E_{\rm rec} = 15.5$--$21.5$~MeV with a single spectrum for a mixture of $^8$Li, $^8$B, and $^9$C, the isotopes with high endpoint energies and yields. Because the isotope fractions are only loosely constrained by SK data~\cite{2016PhRvD..93a2004Z}, they were varied over wide ranges and the average of the steepest and least steep resulting spectra was taken as the nominal spectrum~\cite{2021PhRvD.104l2002A}. 

This treatment was adequate in previous searches, where the atmospheric $\nu$ backgrounds carried large uncertainties: 70\% on the NCQE rate and 30--40\% on the neutron multiplicity~\cite{2021PhRvD.104l2002A,2026ApJ..1005..101A}. With the improved NCQE rejection and the reduced uncertainties (45\% and 10--30\%, respectively), the fit in this energy range becomes more sensitive to the spallation modeling. This is most problematic in the pure-water phases, where most of the signal lies in the $N_{\rm n} \neq 1$ subregion alongside this background. A fit to these phases with the previous threshold indicated an additional contribution at $E_{\rm rec} = 15.5$--$21.5$~MeV, comparable in size to the expected DSNB signal there (see Appendix~B). As this is most likely residual spallation background not fully described by the current modeling, we conservatively raise the $N_{\rm n} \neq 1$ threshold to $E_{\rm rec} = 21.5$~MeV in all phases, above which the spallation contribution is negligible. For the reference DSNB model~\cite{2009PhRvD..79h3013H}, these threshold changes reduce the expected signal by 13.9\% overall. 

\begin{figure}[t]
    \centering
    \includegraphics[width=\columnwidth]{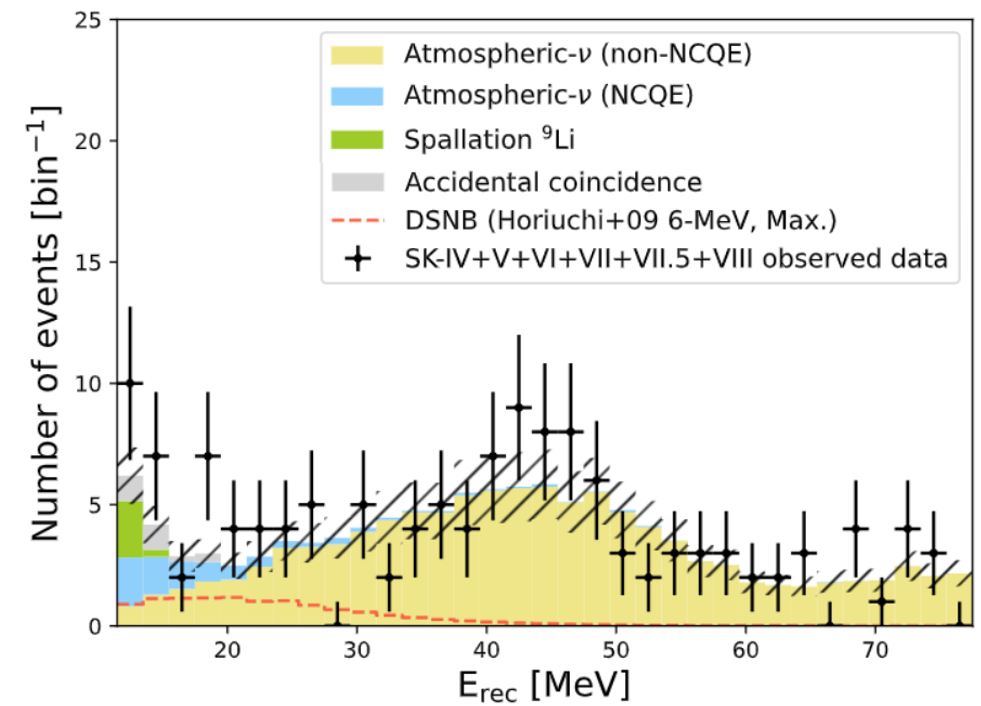}
    \caption{Energy spectra of the data and expected backgrounds in the $N_{\text{n}} = 1$ signal region. The grey hatch represents the total uncertainty on the expected backgrounds. Background-only $p$-values for each 2~MeV bin for $E_{\rm rec} \in [11.5, 29.5]$~MeV are 0.12, 0.14, 0.77, 0.039, 0.24, 0.32, 0.46, 0.27, and 1.00. The nominal Horiuchi+09 DSNB model prediction~\cite{2009PhRvD..79h3013H} is overlaid for reference.}
    \label{fig:binned_final_spectra}
\end{figure}

Atmospheric $\nu$ backgrounds are divided into four categories based on their final state kinematics. CC interactions of $\nu_e$ and $\bar{\nu}_e$ form a ramp spectrum above $\sim$55 MeV (``$\nu_e$ CC''). CC interactions of $\nu_\mu$ and $\bar{\nu}_\mu$ that produce muons below Cherenkov threshold yield Michel electrons, forming a bump spectrum between $\sim$20 and 55 MeV (``Decay $e$''). Events with visible muons or pions tend to be reconstructed at smaller $\theta_{\mathrm{C}}$ (``$\mu/\pi$''), while NCQE and other NC interactions producing nuclear deexcitation $\gamma$ rays tend to be reconstructed at larger $\theta_{\mathrm{C}}$ (``NC$\gamma$''). Background events from the atmospheric $\nu$ flux below $100$~MeV, which is not included in the nominal simulation, were estimated by simulating the fluxes of~\cite{2005APh....23..526B,2020PhRvD.102f3024O} with SKSNSim~\cite{2024ApJ...965...91N} and applying the full event selection. They contribute a few percent of the nominal atmospheric $\nu$ background at the lowest energies, rising to 10\% at $E_{\rm rec} \sim 30$~MeV. Since their spectral shape resembles that of the $\nu_e$ CC component, the normalization uncertainty assigned to that component covers this contribution.

\begin{figure*}[htbp]
    \centering
    \includegraphics[width=\textwidth]{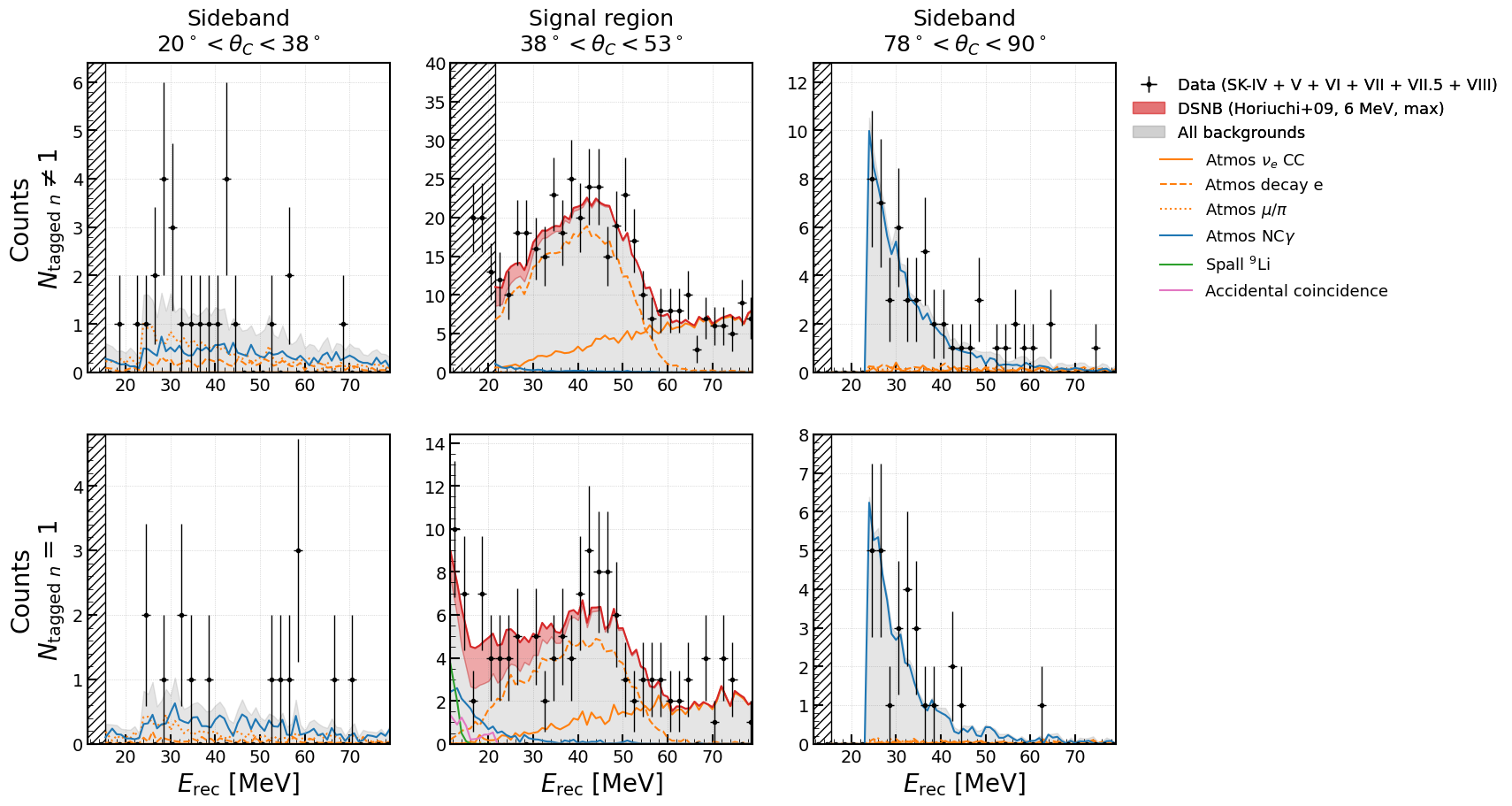}
    \caption{Best fit of the DSNB signal and six background components to the 5003.0-day dataset. The DSNB signal region is the mid-$\theta_{\mathrm{C}}$ band. The signal is stacked above the total background in grey. Data are shown in 2 MeV bins for visualization; the fit itself is unbinned in energy. Data below 21.5 MeV in the middle upper panel are also shown but not used.}
    \label{fig:bestfit_stacked_2mev}
\end{figure*}

An unbinned extended maximum-likelihood fit was performed on the multi-phase dataset, with a single DSNB rate shared across all phases. Parameters describing effects correlated across phases (the atmospheric and spallation background normalizations, along with their nuisance parameters) are shared. Parameters reflecting phase-specific detector conditions (accidental coincidence rates and signal nuisance parameters) are fit per phase.

Systematic uncertainties are incorporated as nuisance parameters of three classes: normalization, migration, and shape systematics. Each background normalization is constrained by a Gaussian penalty centered on its nominal MC prediction, with a width equal to its uncertainty. The DSNB normalization is free to float, but carries an uncertainty of 2.9\% for SK-IV and 3.1\% for SK-V--VIII, combining the IBD cross section, livetime, fiducial volume, and event selection uncertainties in quadrature. 

Migration systematics move events between the $N_{\text{n}}$ regions while preserving the total. All components carry the neutron tagging efficiency uncertainty. Atmospheric backgrounds additionally carry the neutron multiplicity uncertainty, with separate parameters for the CC-like and NC categories, allowing them to migrate in different directions. Since spallation $^{9}$Li and accidental coincidences are fit only in the $N_{\text{n}}=1$ region, the $N_{\text{n}}$ migration systematic instead scales their normalization. Migration of NC$\gamma$ events from high to mid $\theta_{\mathrm{C}}$ due to the $\gamma$ multiplicity uncertainty lies well within the $45$\% normalization uncertainty and is not included.

The only shape uncertainty is that on the spallation $^{9}$Li spectrum. The $30\%$ uncertainty on the $^{9}$Li remaining rate is implemented as an allowed change in the spectral shape above and below $E_{\rm rec} = 13.5$~MeV, preserving the total event rate. The systematics implementation for each fit component is summarized in Table~\ref{tab:tab_2}. 

Sensitivity studies show that the updated systematic uncertainties and spectral fitting method, together with the lowered $N_{\rm n} = 1$ threshold, more than double the median discovery potential (see Appendix~C).

The best-fit configuration from a joint fit of the full 5003.0-day dataset is shown in Figure ~\ref{fig:bestfit_stacked_2mev}. Using the Horiuchi+09 6~MeV model~\cite{2009PhRvD..79h3013H} for the signal shape, the best-fit DSNB flux is $3.6^{+1.6}_{-1.5}~\mathrm{cm^{-2}\,s^{-1}}$ ($4.8^{+2.2}_{-2.0}~\mathrm{events\,yr^{-1}}$) for $E_\nu > 13.3$~MeV. The profile likelihood ratio (Figure~\ref{fig:profile_llh_flux}) has a single well-defined minimum and disfavors the null hypothesis at $2.6\sigma$ ($99.5\%$ C.L., one-sided). Using only the Gd water phases (pure water phases), the flux is determined as $3.9^{+2.2}_{-1.9}~\mathrm{cm^{-2}\,s^{-1}}$ ($3.3^{+2.6}_{-2.4}~\mathrm{cm^{-2}\,s^{-1}}$) from a fitted event rate of $5.2^{+2.9}_{-2.6}~\mathrm{events\,yr^{-1}}$ ($4.4^{+3.5}_{-3.2}~\mathrm{events\,yr^{-1}}$) disfavoring the null hypothesis by $2.2\sigma$ ($1.4\sigma$). Full fit results are provided in Table~\ref{tab:tab_3}. 

\textit{Discussion---}The best-fit DSNB flux lies near the upper end of the $2.1$--$3.9~\mathrm{cm^{-2}\,s^{-1}}$ range predicted by the Horiuchi+09 6~MeV model~\cite{2009PhRvD..79h3013H}. The observed flux using this best-fit value for normalization is shown against the predicted flux in Figure~\ref{fig:dsnb_global_summary}. Repeating the fit with the Kresse+21~\cite{2021ApJ...909..169K} and Ashida+23~\cite{2023ApJ...953..151A} signal spectral shapes gives similar significances ($2.7\sigma$ and $2.6\sigma$, respectively). This analysis combines data from SK-IV~\cite{2021PhRvD.104l2002A} and V (pure water) as well as VI, VII~\cite{2026ApJ..1005..101A}, VII.5, and VIII (Gd water). For $E_\nu>17.3$~MeV, this result corresponds to $1.8^{+0.8}_{-0.7}$~cm$^{-2}$s$^{-1}$, consistent within uncertainties with the $1.4^{+1.5}_{-1.2}$~cm$^{-2}$s$^{-1}$ obtained from~\cite{2026ApJ..1005..101A}.

\textit{Conclusion---}We report the first indication of the DSNB $\bar{\nu}_e$ signal, based on a spectral fit of 5003.0 days of pure water (after 2008) and Gd-water data in SK. The best-fit flux disfavors the background-only hypothesis at $2.6\sigma$ for $E_\nu>13.3$~MeV. Increased Gd-water exposure, improved atmospheric $\nu$ background rejection, and reduced systematic uncertainties allowed the $N_{\text{n}} = 1$ threshold to be lowered to $E_\nu>13.3$~MeV, with the $N_{\text{n}} \neq 1$ threshold conservatively raised to $E_\nu>23.3$~MeV. A complementary DSNB model-independent analysis gives a $1.8\sigma$ tension with the background-only hypothesis for the energy range $13.3 < E_{\nu} < 31.3$~MeV.

\begin{figure}[htbp]
    \centering
    \includegraphics[width=\columnwidth]{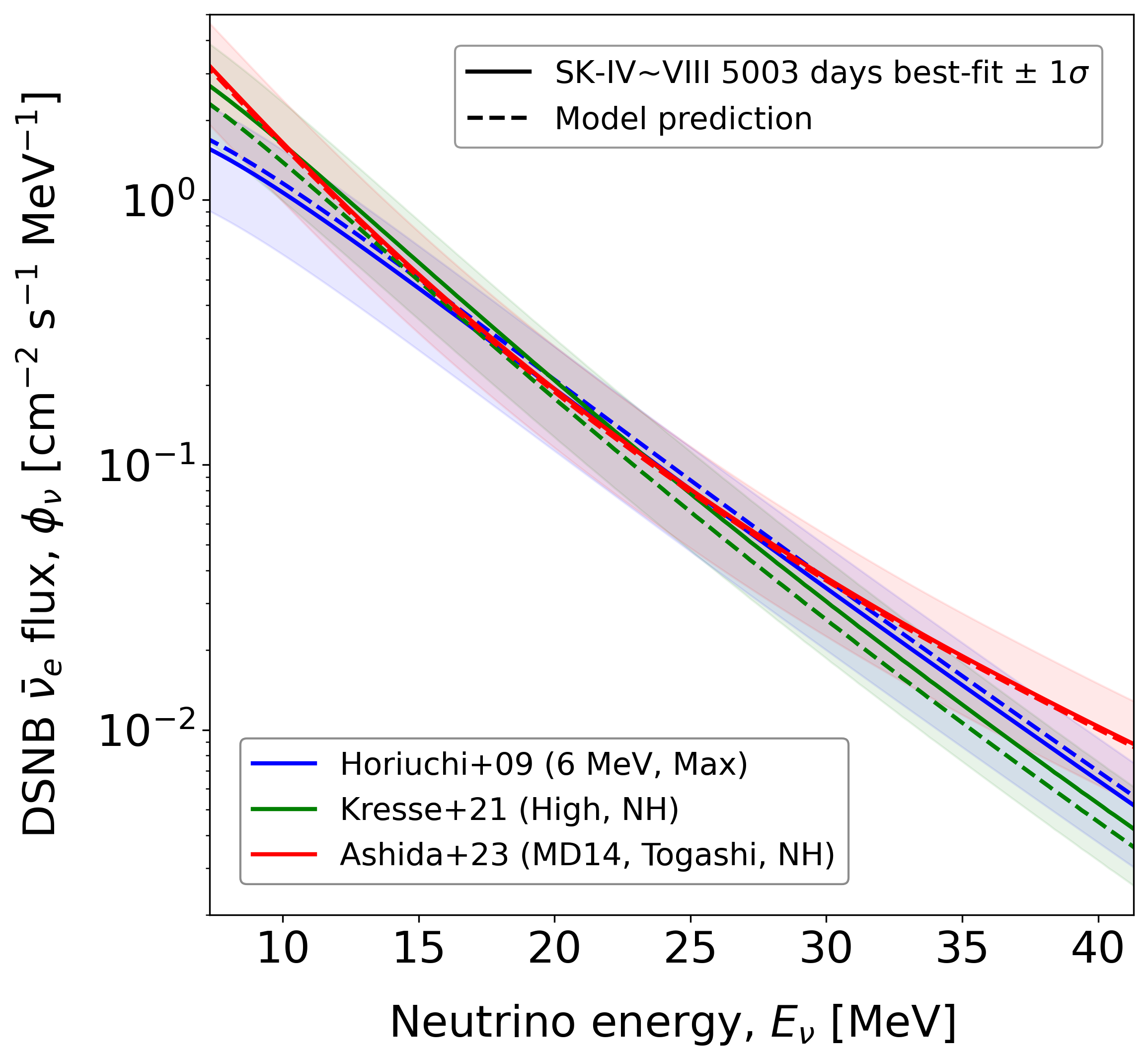}
    \caption{The best-fit $\pm 1\sigma$ normalization of the DSNB flux spectra from Horiuchi+09~\cite{2009PhRvD..79h3013H}, Kresse+21~\cite{2021ApJ...909..169K}, and Ashida+23~\cite{2023ApJ...953..151A} models are shown (solid line and shaded area) along with the original model predictions (dashed line).}
    \label{fig:dsnb_global_summary}
\end{figure}

\textit{Acknowledgements---}
We gratefully acknowledge the cooperation of the Kamioka Mining and Smelting Company. The Super-Kamiokande experiment has been built and operated from funding by the Japanese Ministry of Education, Culture, Sports, Science and Technology; the U.S. Department of Energy; and the U.S. National Science Foundation. Some of us have been supported by funds from the National Research Foundation of Korea (NRF-2009-0083526, NRF-2022R1A5A1030700, NRF-2022R1A3B1078756, RS-2025-00514948, RS-2025-23525600, RS-2026-25489835) funded by the Ministry of Science, Information and Communication Technology (ICT); the Institute for Basic Science (IBS-R016-D1); and the Ministry of Education (2018R1D1A1B07049158, 2021R1I1A1A01042256, RS-2024-00442775); the Japan Society for the Promotion of Science; the National Natural Science Foundation of China (Grants No. 12375100 and 12521007); the Spanish Ministry of Science, Universities and Innovation (grant PID2024-157541NB-I00); the Natural Sciences and Engineering Research Council (NSERC) of Canada; the Scinet and Digital Research of Alliance Canada; the National Science Centre (UMO-2022/46/E/ST2/00336) and the Ministry of  Science and Higher Education (2023/WK/04), Poland; the Science and Technology Facilities Council (STFC) and Grid for Particle Physics (GridPP), UK; the European Union's Horizon Europe Research and Innovation Programme HORIZON-MSCA-SE-2023 JENNIFER3 project, grant agreement no.101183137, H2020-MSCA-RISE-2019 SK2HK grant agreement no. 872549; and European Union's Next Generation EU/PRTR  grant CA3/RSUE2021-00559; the National Institute for Nuclear Physics (INFN), Italy; Japan Society for the Promotion of Science (JSPS) Grant-in-Aid no.~25K24689, 26K00707, and 26K17172.

\bibliography{apssamp} 

\label{sec:endmatter}

\clearpage
\onecolumngrid
\begin{center}
  \textbf{\large End Matter}
\end{center}
\twocolumngrid

\onecolumngrid
\begin{center}
  \includegraphics[width=\textwidth]{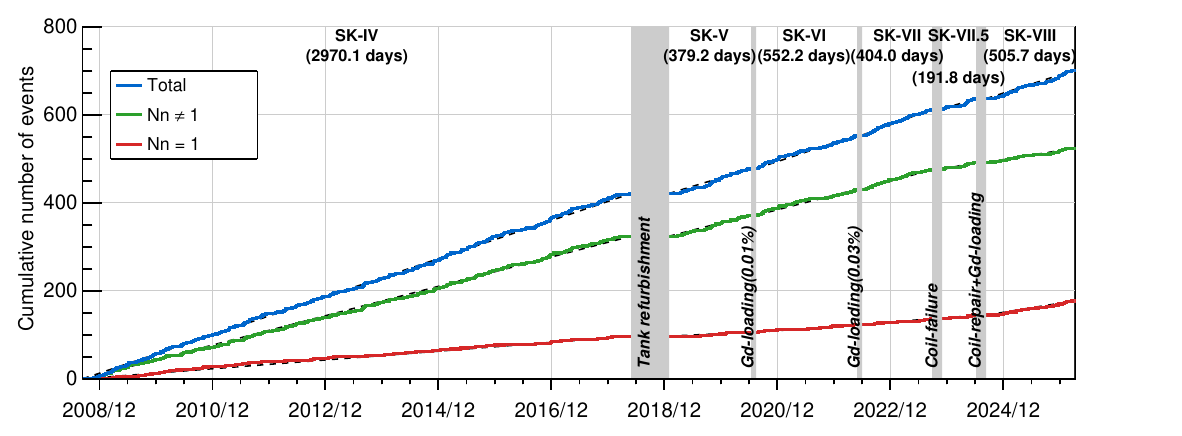}
  \captionof{figure}{Number of events accumulated over the six data-taking phases used in this analysis. The blue line shows the total number of events, while the red (green) line shows events with the number of tagged neutrons, $N_{\rm n} = 1$ ($N_{\rm n} \neq 1$).}
  \label{fig:data_taking}
\end{center}
\twocolumngrid

\textit{Appendix A: Data taking history---}Figure~\ref{fig:data_taking} shows the accumulation of events in the final data sample over the six phases covered in our analysis. The $N_\mathrm{n}=1$ rate increases after each Gd loading owing to the higher neutron tagging efficiency.

\textit{Appendix B: $N_{\rm n} \neq 1$ threshold and signal impact---}For the pure-water phases, keeping the $N_{\rm n} \neq 1$ threshold at the previous value of $E_{\rm rec} = 15.5$~MeV with a spallation model based on that of previous searches~\cite{2021PhRvD.104l2002A,2026ApJ..1005..101A} increases the fitted DSNB rate by about its $1\sigma$ uncertainty. Because the fits with the previous and current thresholds share all data except the $N_{\rm n} \neq 1$ events in $E_{\rm rec} = 15.5$--$21.5$~MeV, a shift of this size is larger than expected from statistical fluctuations alone. Since spallation decays without a neutron enter the $N_{\rm n} \neq 1$ subregion regardless of the Gd concentration, we apply the higher threshold to all phases. Excluding this energy range lowers the expected signal and is the conservative choice.

For the reference DSNB model~\cite{2009PhRvD..79h3013H}, including the energy-dependent signal efficiencies, lowering the $N_{\text{n}} = 1$ threshold increases the livetime-averaged expected signal in that subregion by 23.3\%, while raising the $N_{\text{n}} \neq 1$ threshold reduces it in the other by 41.0\%. The net effect is a signal loss of 24.0\% in the pure-water phases and 3.2\% in the 0.011\% Gd-water phase, and a signal gain of 10.7\% in the 0.033\% Gd-water phases, corresponding to a loss of 13.9\% across the entire dataset. The fraction of the expected signal in the higher-purity $N_{\rm n} = 1$ subregion increases from 42\% with a common threshold of $E_{\rm rec} = 15.5$~MeV to 60\% with the new thresholds. Other signal models give similar values. 

\textit{Appendix C: Sensitivity study---}Sensitivity studies were performed with pseudo-datasets corresponding to 4040 days of 0.033\% Gd-water livetime, ten times the SK-VII exposure, in which the Horiuchi+09 6~MeV signal prediction was injected along with the expected backgrounds. The combination of the normalization penalty terms, the updated NCQE, spallation $^{9}$Li, and neutron multiplicity systematics, and lowering of the analysis threshold in the $N_{\text{n}} = 1$ signal subregion from $E_{\rm rec} = 15.5$~MeV to $11.5$~MeV more than doubled the expected median discovery potential, with the $N_{\text{n}} \neq 1$ threshold held at $E_{\rm rec} = 21.5$~MeV.

\textit{Appendix D: Profile likelihood---}Figure~\ref{fig:profile_llh_flux} shows $-2\Delta\ln\mathcal{L}$ versus DSNB flux, profiled over all nuisance parameters. 

\begin{figure}[h]
    \centering
    \includegraphics[width=\columnwidth]{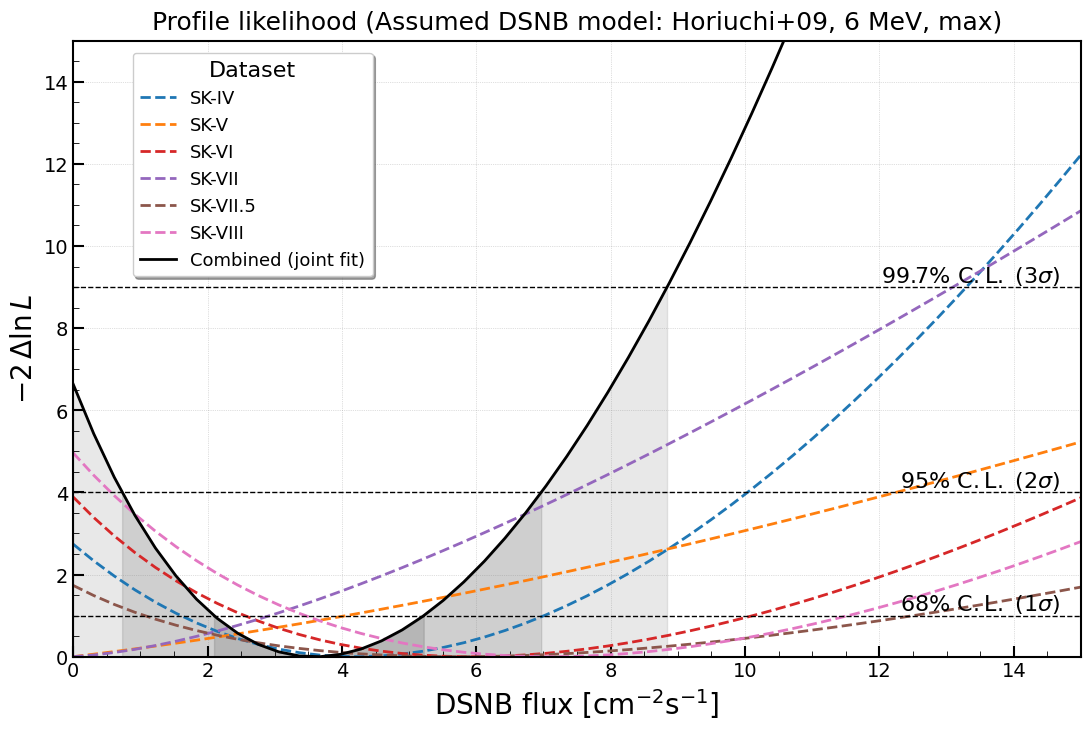}
    \caption{Profile likelihood ratio as a function of the DSNB flux for $E_{\nu} > 13.3$~MeV. The solid black curve represents the joint fit. The colored dashed curves represent the separate per-phase fits, shown only for comparison.}
    \label{fig:profile_llh_flux}
\end{figure}

\begin{table*}
\renewcommand{\arraystretch}{1.3}
\caption{\label{tab:tab_1}Systematic uncertainties considered in the spectral
fit. The labels are referenced in Table~\ref{tab:tab_2}.}
\begin{ruledtabular}
\begin{tabular}{lcccccc}
Item & SK-IV & SK-V & SK-VI & SK-VII & SK-VII.5 & SK-VIII \\
\colrule
(A) Atmospheric $\nu$ flux & $\pm20\%$ & $\pm20\%$ & $\pm20\%$ & $\pm20\%$ & $\pm20\%$ & $\pm20\%$ \\
(B) Neutrino cross section & $\pm30\%$ & $\pm30\%$ & $\pm30\%$ & $\pm30\%$ & $\pm30\%$ & $\pm30\%$ \\
(C) Atmospheric $\nu$ NCQE rate & $\pm45\%$ & $\pm45\%$ & $\pm45\%$ & $\pm45\%$ & $\pm45\%$ & $\pm45\%$ \\
(D) Neutron multiplicity & $^{+27}_{-30}\%$ & $^{+27}_{-30}\%$ & $^{+18}_{-31}\%$ & $^{+12}_{-32}\%$ & $^{+12}_{-32}\%$ & $^{+12}_{-32}\%$ \\
(E) Neutron tagging efficiency & $\pm10\%$ & $\pm10\%$ & $\pm3\%$ & $\pm5\%$ & $\pm5\%$ & $\pm5\%$ \\
(F) Spallation $^9\mathrm{Li}$ yield & $\pm22\%$ & $\pm22\%$ & $\pm22\%$ & $\pm22\%$ & $\pm22\%$ & $\pm22\%$ \\
(G) Spallation $^9\mathrm{Li}$ remaining rate & $\pm30\%$ & $\pm30\%$ & $\pm30\%$ & $\pm30\%$ & $\pm30\%$ & $\pm30\%$ \\
(H) Accidental coincidence rate & $\pm5\%$ & $\pm5\%$ & $\pm5\%$ & $\pm5\%$ & $\pm5\%$ & $\pm5\%$ \\
\end{tabular}
\end{ruledtabular}
\end{table*}

\begin{table*}[t]
\renewcommand{\arraystretch}{1.5}
\caption{Signal and background components of the spectral fit and their associated systematics. Entries in parentheses refer to the items in Table~\ref{tab:tab_1}; where several are listed, they are combined in quadrature. ``Normalization'' lists the items constraining the overall normalization of each component, which is free to float for the DSNB. ``Migration'' moves events between the $N_{\mathrm{n}}$ regions, preserving the total; for spallation $^{9}$Li and accidental coincidences, which are fit only in the $N_{\mathrm{n}} = 1$ signal subregion, the neutron tagging uncertainty acts as a pure scaling and is listed under ``Normalization'' instead. ``Shape'' gives the spectral-shape uncertainty, implemented as described in the text.}
\label{tab:tab_2}\begin{ruledtabular}
\begin{tabular}{lcccc}
Component & Fitted Region & Normalization & Migration & Shape \\
\colrule
DSNB & Mid-$\theta_{\mathrm{C}}$, both $N_{n}$ & Free-floating & (E) & --- \\
Atmospheric $\nu_e~\mathrm{CC}$ & Mid-$\theta_{\mathrm{C}}$, both $N_{n}$ & (A), (B) & (D), (E) & --- \\
Atmospheric $\mathrm{Decay}\text{-}e$ & Mid-$\theta_{\mathrm{C}}$, both $N_{n}$ & (A), (B) & (D), (E) & --- \\
Atmospheric $\mu/\pi$ & Low-$\theta_{\mathrm{C}}$, both $N_{n}$ & (A), (B) & (D), (E) & --- \\
Atmospheric $\mathrm{NC}\gamma$ & Mid/High-$\theta_{\mathrm{C}}$, both $N_{n}$ & (C) & (D), (E) & --- \\
Spallation $^9\mathrm{Li}$ & Mid-$\theta_{\mathrm{C}}$, $N_{n}=1$ & (E), (F) & --- & (G) \\
Accidental coincidence & Mid-$\theta_{\mathrm{C}}$, $N_{n}=1$ & (E), (H) & --- & --- \\
\end{tabular}
\end{ruledtabular}
\end{table*}

\begin{table*}[t]
\renewcommand{\arraystretch}{1.5}
\caption{\label{tab:tab_3}%
Parameters of the spectral fit and their best-fit values from the joint fit and from the separate per-phase fits. Signal and background normalizations ($N$) are shared across phases in the joint fit, except for the accidental coincidences which are constrained independently in each phase. Correlated backgrounds are expressed as scale factors relative to the nominal MC expectation, taken as the central value for the penalty term. Nuisance parameters ($\epsilon$) are pulls in units of their systematic uncertainty.}
\begin{ruledtabular}
\begin{tabular}{lllccccccc}
 Component & Parameter & Type & Joint & SK-IV & SK-V & SK-VI & SK-VII & SK-VII.5 & SK-VIII \\
\colrule
DSNB & $N_s$ & rate per year & 4.8 & 5.6 & 0.0 & 7.9 & 0.0 & 7.6 & 9.5 \\
Atmospheric $\nu_e~\mathrm{CC}$ & $N_{\nu_e\mathrm{CC}}$ & scale on MC nominal & 1.1 & 1.3 & 1.3 & 0.8 & 0.8 & 1.0 & 0.8 \\
Atmospheric $\mathrm{Decay}\text{-}e$ & $N_{\mathrm{Decay}\text{-}e}$ & scale on MC nominal & 1.2 & 1.2 & 1.2 & 1.2 & 1.3 & 1.2 & 1.0 \\
Atmospheric $\mu/\pi$ & $N_{\mu/\pi}$ & scale on MC nominal & 1.2 & 1.3 & 1.0 & 1.0 & 0.9 & 1.1 & 1.1 \\
Atmospheric $\mathrm{NC}\gamma$ & $N_{\mathrm{NC}\gamma}$ & scale on MC nominal & 1.1 & 1.0 & 1.2 & 1.1 & 1.3 & 0.9 & 0.9 \\
\shortstack[l]{Spallation $^9\mathrm{Li}$}
 & $N_{^9\mathrm{Li}}$ & scale on MC nominal & 1.1 & 1.0 & 1.0 & 1.0 & 1.0 & 1.0 & 1.0 \\
\shortstack[l]{Accidental coincidence}
 & $N_{\mathrm{acc}}$ & rate per phase & --- & 0.2 & 0.5 & 0.2 & 0.3 & 0.3 & 0.1 \\
 \colrule
Atmospheric CC $N_{n}$ migration & $\epsilon_{\mathrm{ntag~CC}}$ & pull in units of $\sigma$ & $-0.2$ & 0.0 & $-1.1$ & $-1.3$ & $-0.9$ & 0.0 & 1.1 \\
Atmospheric NC $N_{n}$ migration & $\epsilon_{\mathrm{ntag~NC}}$ & pull in units of $\sigma$ & 0.8 & 0.2 & 0.7 & 0.3 & 0.4 & 0.0 & 0.2 \\
DSNB $N_{n}$ migration & $\epsilon_{\mathrm{ntag~sig}}$ & pull in units of $\sigma$ & --- & 0.2 & 0.0 & 0.1 & 0.0 & 0.0 & 0.2 \\
DSNB signal efficiency & $\epsilon_{\mathrm{eff~sig}}$  & pull in units of $\sigma$ & --- & 0.0 & 0.0 & 0.0 & 0.0 & 0.0 & 0.0 \\
$^9\mathrm{Li}$ spectral shape & $\epsilon_{^9\mathrm{Li}}$ & pull in units of $\sigma$ & 0.1 & 0.1 & 0.0 & 0.0 & 0.0 & 0.0 & $-0.1$ \\
\end{tabular}
\end{ruledtabular}
\end{table*}

\clearpage
\setcounter{figure}{0}

\onecolumngrid
\begin{center}
  \textbf{\large Supplemental Material for:\\Indication of Electron Antineutrinos from the Diffuse Supernova Neutrino Background}
\end{center}
\twocolumngrid




\renewcommand{\thefigure}{S\arabic{figure}}
\renewcommand{\thetable}{S\arabic{table}}
\renewcommand{\theequation}{S\arabic{equation}}
\renewcommand{\thesection}{S\arabic{section}}
\renewcommand{\bibnumfmt}[1]{[S#1]}
\renewcommand{\citenumfont}[1]{S#1}



\preprint{APS/123-QED}

\title{Supplemental Material for:\\Indication of Electron Antineutrinos from the Diffuse Supernova Neutrino Background}


\author{}
\collaboration{Super-Kamiokande Collaboration}


\maketitle

This Supplemental Material shows the energy spectra of Figs.~1 and 2 of the main text separately for the pure-water phases (SK-IV and SK-V) and the Gd-loaded phases (SK-VI, SK-VII, SK-VII.5, and SK-VIII). Figures~S1 and S2 correspond to Fig.~1, and Figs.~S3 and S4 to Fig.~2.
\newpage


\begin{figure}[h]
\includegraphics[width=0.5\textwidth]{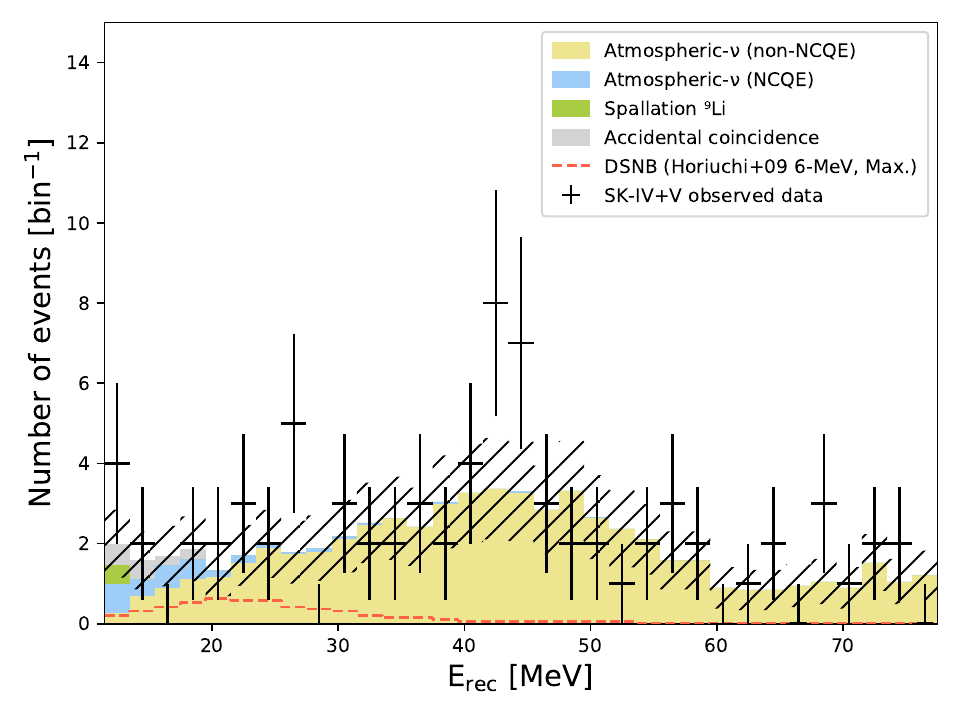}
\caption{Energy spectra of the data and expected backgrounds in the $N_{\rm n} = 1$ signal region for the pure-water phases (SK-IV and SK-V, 3349.3 days), corresponding to Fig.~1 of the main text. The grey hatch represents the total uncertainty on the expected backgrounds. The nominal Horiuchi+09 DSNB model prediction is overlaid for reference.}
\label{fig:binned_pure_water}
\end{figure}

\begin{figure}[h]
\includegraphics[width=0.5\textwidth]{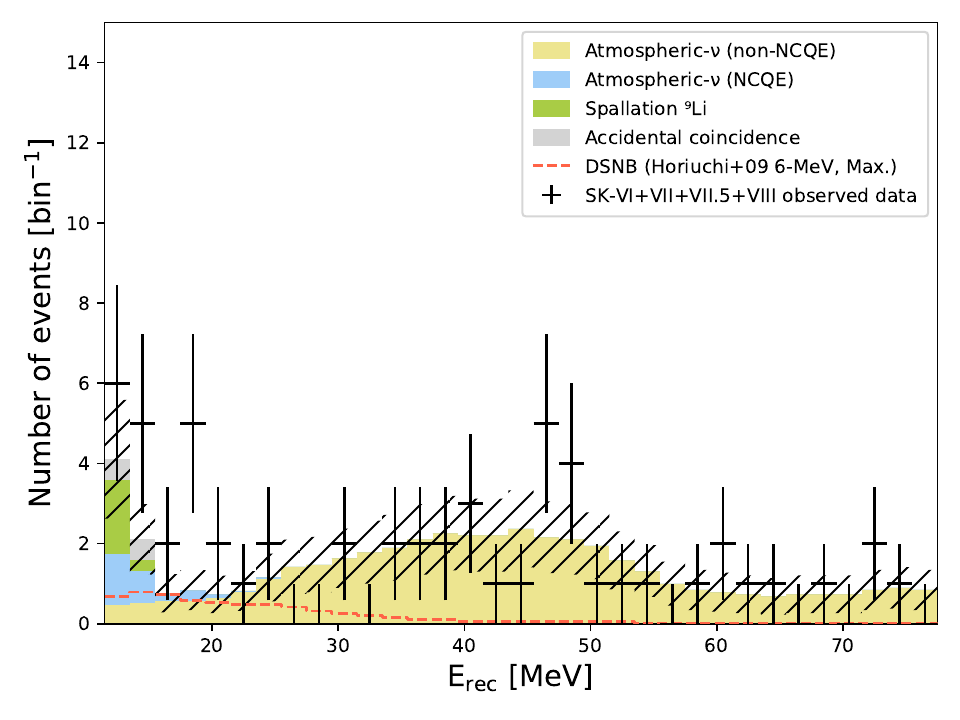}
\caption{Same as Fig.~S1, for the Gd-loaded phases (SK-VI, SK-VII, SK-VII.5, and SK-VIII, 1653.7 days).}
\label{fig:binned_gd_water}
\end{figure}


\begin{figure*}[h]
\includegraphics[width=\textwidth]{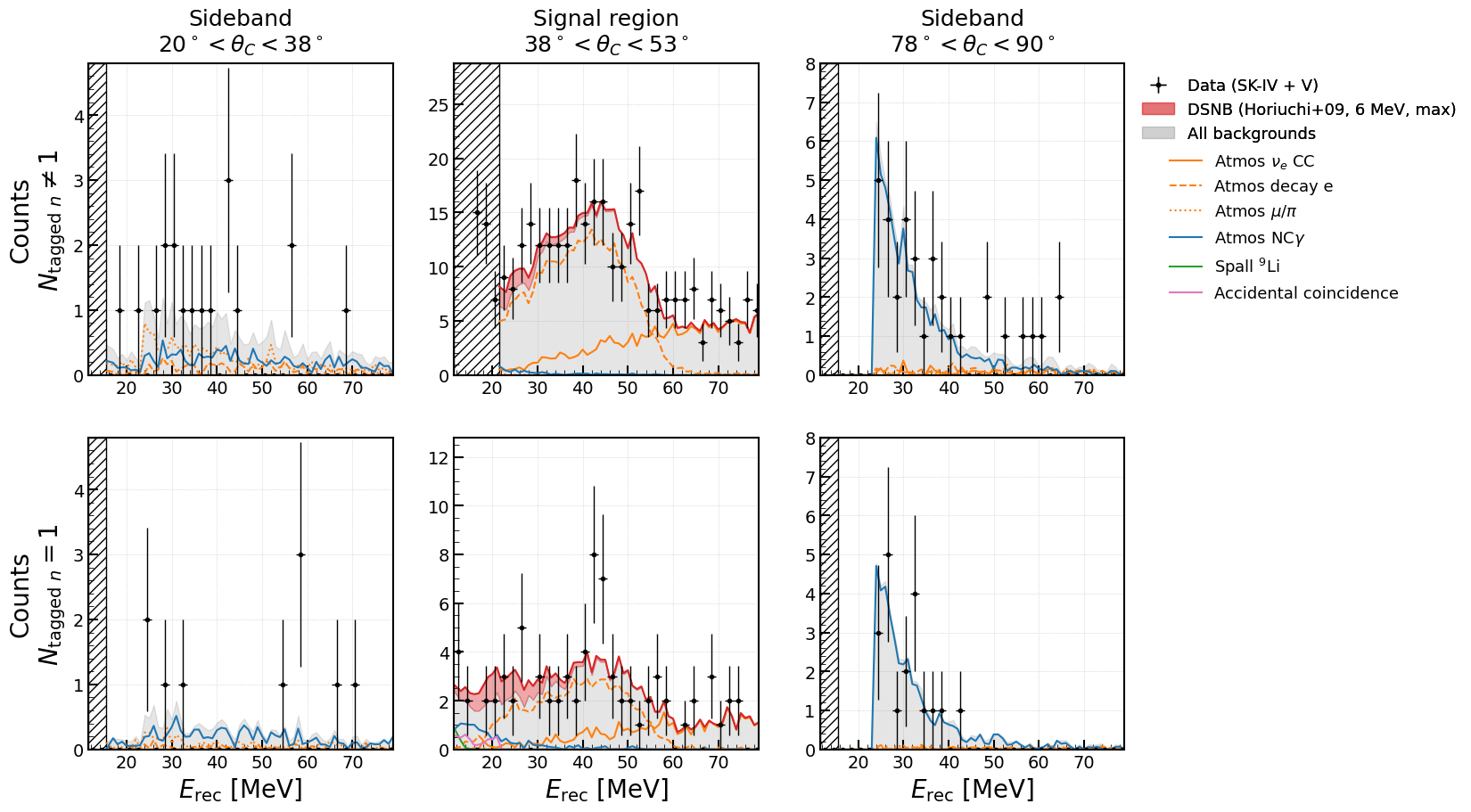}
\caption{Joint-fit prediction from Fig.~2 of the main text, shown separately for the pure-water phases (SK-IV and SK-V, 3349.3 days). The predictions are evaluated at the best-fit parameters of the joint fit to all phases, not from a separate fit to these phases. The DSNB signal is stacked above the total background in grey. Data are shown in 2 MeV bins for visualization; the fit itself is unbinned in energy. Hatched regions lie outside the fit range; data below 21.5 MeV in the middle upper panel are shown but not used.}
\label{fig:spec_pure_water}
\end{figure*}

\begin{figure*}[h]
\includegraphics[width=\textwidth]{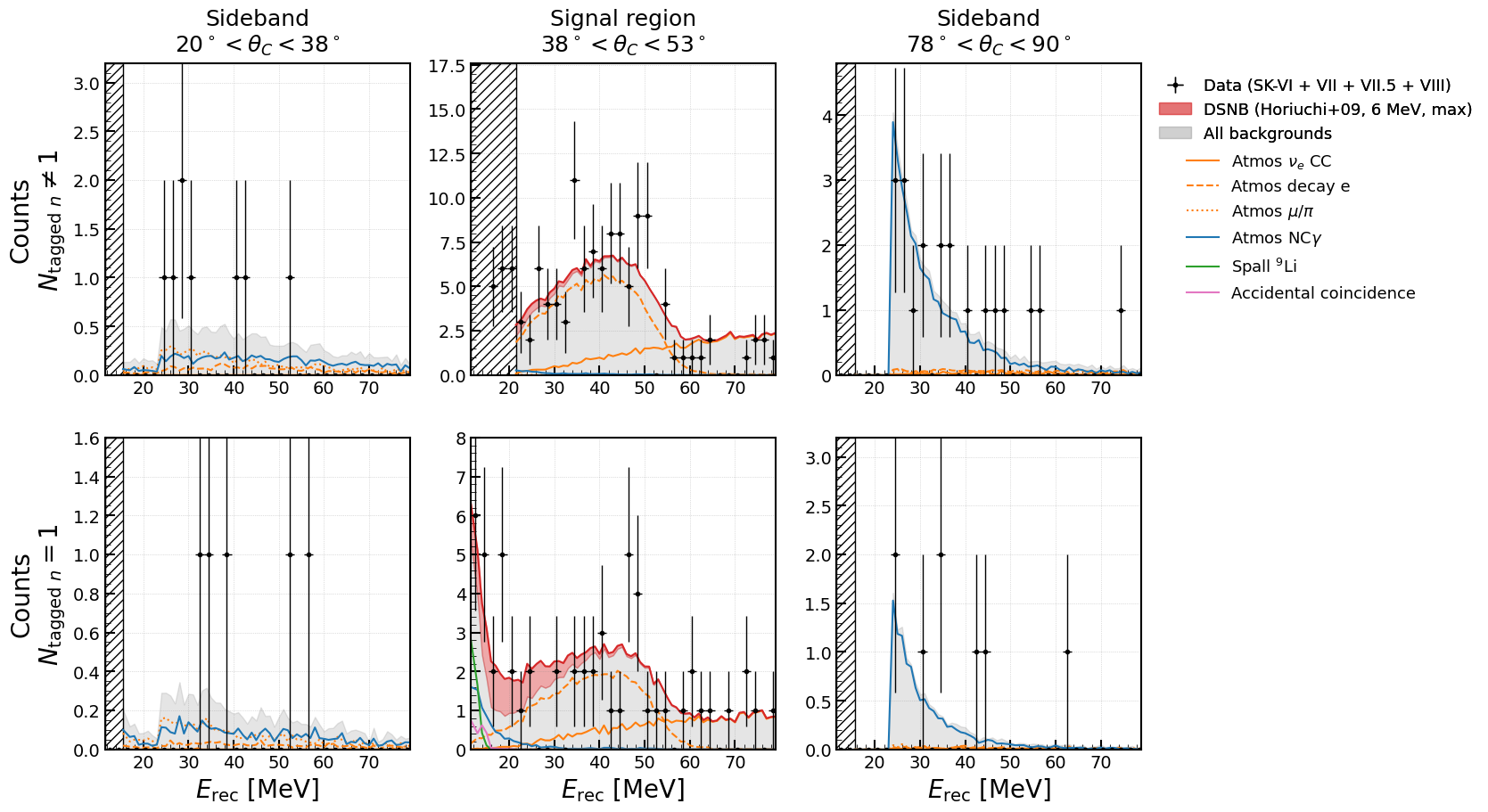}
\caption{Same as Fig.~S3, for the Gd-loaded phases (SK-VI, SK-VII, SK-VII.5, and SK-VIII, 1653.7 days).}
\label{fig:spec_gd_water}
\end{figure*}

%

\end{document}